\documentclass[aps,prd,12pt,nofootinbib,superscriptaddress]{revtex4-2}
\usepackage{bm}
\usepackage{ulem}
\usepackage{comment}
\usepackage{amsmath}
\usepackage{amssymb}
\usepackage{booktabs}
\usepackage{graphicx}
\usepackage{float}
\usepackage{comment}
\usepackage{hyperref}
\usepackage{CJK}
\usepackage{color}
\hypersetup{colorlinks=true, linkcolor=red, citecolor=blue}

\begin{document}
\begin{CJK*}{UTF8}{gbsn}
\title{Null tests with Gaussian Process}

\author{Shengqing Gao}
\email{gaoshengqing@hust.edu.cn}
\affiliation{School of Physics, Huazhong University of Science and Technology, 1037 LuoYu Rd, Wuhan, Hubei 430074, China}
%\author{Qing Gao (郜青)}
\author{Qing Gao}
\email{Corresponding author. gaoqing1024@swu.edu.cn}
\affiliation{School of Physical Science and Technology, Southwest University, Chongqing 400715, China}

\author{Yungui Gong}
%\author{Yungui Gong (龚云贵)}
\email{yggong@hust.edu.cn}
\affiliation{Institute of Fundamental Physics and Quantum Technology, Department of Physics, School of Physical Science and Technology,\\
Ningbo University, Ningbo, Zhejiang 315211, China}
\affiliation{School of Physics, Huazhong University of Science and Technology, 1037 LuoYu Rd, Wuhan, Hubei 430074, China}
\author{Xuchen Lu}
\email{Luxc@hust.edu.cn}
\affiliation{School of Physics, Huazhong University of Science and Technology, 1037 LuoYu Rd, Wuhan, Hubei 430074, China}

\begin{abstract}
We investigate the null tests of spatial flatness and the flat $\Lambda$CDM model using the Baryon Acoustic Oscillation (BAO) data measured by the Dark Energy Spectroscopic Instrument (DESI), the cosmic chronometers (CCH) $H(z)$ data, and the Union3 and Pantheon Plus type Ia supernovae (SNe Ia) datasets.
We propose a novel non-parametric reconstruction of $F_{AP}$, $D_M/r_d$ and $D'_M/r_d$ from the DESI BAO data to perform the $Ok$ diagnostic,
and we also conduct the $Ok$ diagnostic using the combination of CCH and SNe Ia data.
The novel method avoids the issue of the dependence on cosmological parameters such as the value of the Hubble constant.
There is no evidence of deviation from the flat $\Lambda$CDM model, nor is there any indication of dynamical dark energy found in the observational data. Since we employ a non-parametric reconstruction method, all the conclusions drawn in this paper remain robust and agnostic to any cosmological model and gravitational theory.
\end{abstract}

%\keywords{null test, Gaussian Process, dark energy, observational data}

\maketitle
\end{CJK*}

\section{Introduction}

The discovery that the Universe is apparently in accelerating expansion, as observed through type Ia supernovae (SNe Ia), is arguably one of the most significant achievements in modern cosmology \cite{SupernovaSearchTeam:1998fmf,SupernovaCosmologyProject:1998vns}.
To account for the current accelerated expansion, numerous models, such as dark energy or modifications of general relativity, have been proposed \cite{Ratra:1987rm,Wetterich:1987fm,Caldwell:1997ii,Zlatev:1998tr,Steinhardt:1999nw,Dvali:2000hr,Carroll:2003wy,Nojiri:2003ft,Starobinsky:2007hu,Hu:2007nk,deRham:2010kj,Gong:2012yv}; for reviews, please see Refs. \cite{Sahni:1999gb,Copeland:2006wr,Padmanabhan:2007xy,Li:2011sd,Benetti:2019gmo}.
One of the popular explanations is the $\Lambda$CDM model, which includes  $5\%$ baryonic matter, $25\%$ cold dark matter (CDM), and $70\%$
cosmological constant, as well as small contributions from massive neutrinos and radiation, and it aligns excellently with observational data. However, aside from the fine-tuning and coincidence problems faced by
the $\Lambda$CDM model,
the theoretical estimation of vacuum energy exceeds astronomical measurements by many orders of magnitude \cite{Weinberg:1988cp}.
Due to the challenges encountered by the $\Lambda$CDM model, the nature of dark energy, particularly whether it can be explained by the cosmological constant, remains one of the biggest unresolved problems in modern cosmology.

The results of the measurements of baryon acoustic oscillations (BAO) in the recent Data Release 1 (DR1) from the first year of observations by the Dark Energy Spectroscopic Instrument (DESI) suggest the presence of dynamical dark energy \cite{DESI:2024mwx}.
Specifically, employing the Chevallier-Polarski-Linder (CPL) parametrization \cite{Chevallier:2000qy,Linder:2002et} for the equation of state of dark energy,
the combination of DESI BAO data and the CMB data measured by {\it Planck} \cite{Planck:2018vyg}
gives $w_0=-0.45^{+0.34}_{-0.21}$
and $w_a=-1.79^{+0.48}_{-1.0}$,
indicating a preference for dynamical dark energy at the $\sim 2.6\sigma$ significance level \cite{DESI:2024mwx}.
The evidence for dynamical dark energy provided by the DESI BAO data has sparked a lot of discussions \cite{Giare:2024gpk,Chan-GyungPark:2024spk,Dinda:2024kjf,Dinda:2024ktd,Jiang:2024xnu,Ghosh:2024kyd,Cortes:2024lgw,Shlivko:2024llw,deCruzPerez:2024shj,Roy:2024kni,Chatrchyan:2024xjj,Perivolaropoulos:2024yxv,Linder:2024rdj,Payeur:2024dnq,Chan-GyungPark:2024brx,Carloni:2024zpl,Gialamas:2024lyw,Luongo:2024fww,RoyChoudhury:2024wri,Wolf:2024eph,Wolf:2024stt,Dias:2024tpf,Sapone:2024ltl,Lu:2024hvv,Gao:2024ily}.
Notably, the model dependence of this dynamical evidence necessitates further examination.
Thus, it is important to explore whether dark energy is indeed dynamical using model-independent methods, including both parametric and non-parametric approaches \cite{Visser:1997qk,Visser:1997tq,Santos:2006ja,Santos:2007pp,Gong:2007fm,Gong:2007zf,Gong:2006tx,Gong:2006gs,Seikel:2007pk,Clarkson:2007pz,Sahni:2008xx,Zunckel:2008ti,Nesseris:2010ep,Clarkson:2010bm,Holsclaw:2010nb,Holsclaw:2010sk,Holsclaw:2011wi,Bilicki:2012ub,Seikel:2012uu,Seikel:2012cs,Shafieloo:2012rs,Shafieloo:2012ht,Gao:2012ef,Gong:2013bn,Yahya:2013xma,Sahni:2014ooa,Nesseris:2014mfa,Cai:2015pia,Li:2015nta,Vitenti:2015aaa,Zhang:2016tto,Wei:2016xti,Yu:2017iju,Yennapureddy:2017vvb,Velten:2017ire,Marra:2017pst,Melia:2018tzi,Gomez-Valent:2018hwc,Pinho:2018unz,Haridasu:2018gqm,Capozziello:2018jya,Arjona:2019fwb,Jesus:2019nnk,Franco:2019wbj,Bengaly:2019ibu,Yang:2019fjt,Dhawan:2021mel,Gangopadhyay:2023nli,Sharma:2024mtq,Li:2014yza,Jiao:2022aep}.
The Gaussian Process (GP) method is a robust statistical technique for modeling distributions over functions within a Bayesian framework \cite{Holsclaw:2010nb,Holsclaw:2010sk,Holsclaw:2011wi,Bilicki:2012ub,Seikel:2012uu,Seikel:2012cs}.
As one of the most widely used non-parametric approaches,
it enables the reconstruction of continuous functions and their derivatives that best represent discrete data points without the need for a predefined functional form.
This flexibility makes GP particularly well-suited for handling complex, nonlinear, and noisy data \cite{Sabogal:2024qxs}.
By assuming Gaussian distributions and employing a covariance (or kernel) function, the GP method effectively characterizes the relationships between data points.
Additionally, its ability to quantify uncertainty alongside predictions makes the GP an invaluable tool in cosmology
\cite{Holsclaw:2010nb,Holsclaw:2010sk,Holsclaw:2011wi,Bilicki:2012ub,Seikel:2012uu,Seikel:2012cs,Seikel:2013fda,Shafieloo:2012ht,Nair:2013sna,Busti:2014dua,Sahni:2014ooa,Verde:2014qea,Li:2015nta,Vitenti:2015aaa,Wang:2016iij,Yang:2015tzc,Cai:2015pia,Zhang:2016tto,Wei:2016xti,Yu:2017iju,Yennapureddy:2017vvb,Melia:2018tzi,Marra:2017pst,Velten:2017ire,Gomez-Valent:2018hwc,Pinho:2018unz,Haridasu:2018gqm,Jesus:2019nnk,Bengaly:2019ibu,Yang:2019fjt,Zhang:2024ndc,Sabogal:2024qxs,Dinda:2024ktd,Yang:2025kgc,Mukherjee:2025fkf,Johnson:2025blf}.

In this paper, we apply the GP method to reconstruct the DESI BAO data, the cosmic chronometers (CCH) $H(z)$ data, and the Union3 \cite{Rubin:2023ovl}  and Pantheon Plus SNe Ia \cite{Scolnic:2021amr} data,
and conduct null tests on the spatial flatness and the flat $\Lambda$CDM model using the $Ok$ diagnostic \cite{Clarkson:2007pz}, the $Om$ diagnostic \cite{Sahni:2008xx,Shafieloo:2012rs} and the $Lz$ test \cite{Zunckel:2008ti}.
The paper is organized as follows:
In Sec. \ref{null_test}, we introduce the null test method.
In Sec. \ref{bao_null} we perform the null tests with the DESI BAO data.
The null tests combining CCH and SNe Ia data are carried out in Sec. \ref{snh_null}.
The conclusion is drawn in Sec. \ref{conclusion}.

\section{Null test Methods}
\label{null_test}
In this section, we review model-independent null test methods which are used to detect possible deviations from spatial flatness and the flat $\Lambda$CDM model.

Assuming the Universe is homogeneous and isotropic on large scales,
the spacetime geometry is described by the Friedmann-Robertson-Walker metric, and the transverse comoving distance $D_{M}$ is
\begin{equation}
\label{eq:D_M}
    D_{M}(z)=\frac{1}{H_0 \sqrt{|\Omega_{k0}|}} \text{sinn} \left[\sqrt{|\Omega_{k0}|} \int_0^z \frac{d x}{E(x)}\right],
\end{equation}
where $H_0=H(z=0)$ is the Hubble constant, $E(z)=H(z)/H_0$ is the dimensionless Hubble parameter,
and $\text{sinn}(\sqrt{|\Omega_{k0}|}x)/\sqrt{|\Omega_{k0}|}=\sinh(\sqrt{|\Omega_{k0}|}x)/\sqrt{|\Omega_{k0}|}$, $x$,
and $\sin(\sqrt{|\Omega_{k0}|}x)/\sqrt{|\Omega_{k0}|}$ for $\Omega_{k0}>0$,
$\Omega_{k0}=0$ and $\Omega_{k0}<0$, respectively.
For a spatially flat universe, $\Omega_{k0}=0$,
\begin{equation}
\label{eq:D_M_flat}
D_M(z)=\frac{1}{H_0}\int_{0}^{z}\frac{d x}{E(x)}.
\end{equation}

\subsection{$Ok$ diagnostic}
From Eq. \eqref{eq:D_M}, we get \cite{Clarkson:2007pz}
\begin{equation}
\label{omkeq}
\Omega_{k0}\equiv -\frac{k}{H_0^2}
%=\frac{[H(z)D_M'(z)]^2 -1}{[H_0 D_M(z)]^2}
=\frac{[E(z)D'(z)]^2 -1}{[D(z)]^2},
\end{equation}
where $D(z)=H_0 D_M(z)$ and $D'(z)=dD(z)/dz$.
The null test of constant curvature or the cosmological principle can be obtained by calculating \cite{Clarkson:2007pz}
\begin{equation}
\begin{split}
\label{omkczeq}
\mathcal{C}(z)&=1+H^2 (D_M D_M''-D_M'^2)+HH'D_M D'_M\\
&=1+E^2 (D D''-D'^2)+EE'D D',
\end{split}
\end{equation}
$\mathcal{C}(z)$ is zero at all redshift if the Friedmann-Robertson-Walker metric is assumed.

Since the denominator in Eq. \eqref{omkeq} is zero at $z=0$, to avoid the problem, we can alternatively utilize the $Ok$ diagnostic \cite{Cai:2015pia}
\begin{equation}
\label{ok1}
\mathcal{O}_k(z)=E(z)D'(z)-1=H(z)D'_M(z)-1
\end{equation}
to test the flatness of the spatial geometry.
For a spatially flat universe, from Eq. \eqref{eq:D_M_flat} we know that $D'(z)=E^{-1}(z)$ and $D'_M(z)=H^{-1}(z)$,
so $\mathcal{O}_k(z)=0$,
and we can reconstruct $\mathcal{O}_k(z)$ from observational data to perform a null test of the Universe's flatness by examining whether $\mathcal{O}_k(z)$ deviates from zero.
Note that the null test \eqref{ok1} is independent of cosmological models and even gravitational theories,
it just depends on the Friedmann-Robertson-Walker metric which is based on the cosmological principle.

\subsection{$Om$ diagnostic}

The two-point $Om$ diagnostic is \cite{Shafieloo:2012rs}
\begin{equation}
\label{om2}
Om(z_2,z_1)=\frac{E^2(z_2)-E^2(z_1)}{(1+z_2)^3-(1+z_1)^3}.
\end{equation}
For the flat $\Lambda$CDM model, $Om(z_2,z_1)=\Omega_{m0}$,
so if the reconstructed $Om(z_2,z_1)$ is a constant, then we conclude that the flat $\Lambda$CDM model is consistent with the observational data used for the reconstruction.
Take $z_1=0$, then we get the $Om$ diagnostic \cite{Sahni:2008xx,Shafieloo:2012rs}
\begin{equation}
\label{om1}
Om(z)=\frac{E^2(z)-1}{(1+z)^3-1}.
\end{equation}
Again $Om(z)=\Omega_{m0}$ for the flat $\Lambda$CDM model.

Instead of $E(z)$, we can replace it with $H(z)$ in Eq. \eqref{om2} and the two-point $Om$ diagnostic becomes
\begin{equation}
\label{om2a}
Om(z_2,z_1)=\frac{H^2(z_2)-H^2(z_1)}{(1+z_2)^3-(1+z_1)^3}.
\end{equation}
For the flat $\Lambda$CDM model, $Om(z_2,z_1)=\Omega_{m0} H_0^2$.
Note that $Om$ diagnostics is a null test for $\Lambda$CDM model.

\subsection{$Lz$ test}
The $Lz$ null test on flat $\Lambda$CDM model is defined as
\cite{Zunckel:2008ti}
\begin{equation}
\label{lz1}
\mathcal{L}(z)=2[(1+z)^3-1]D''(z)+3(1+z)^2D'(z)[1-D'(z)^2],
\end{equation}
$\mathcal{L}(z)=0$ for the flat $\Lambda$CDM model, so by reconstructing $\mathcal{L}(z)$ from observational data, we can determine whether
$\mathcal{L}(z)=0$ for the flat $\Lambda$CDM model, so by reconstructing $\mathcal{L}(z)$ from observational data, we can determine whether the flat $\Lambda$CDM model is consistent with the observational data.

\section{DESI BAO data}
\label{bao_null}
DESI BAO dataset includes the determinations of $D_V(z)/r_d$ at two effective redshifts $z_\text{eff}=0.30$
and $z_\text{eff}=1.48$,
the measurements of $D_M(z)/r_d$ and
$D_H(z)/r_d$ at five different redshits, where $D_H(z)=1/H(z)$ and
the sound horizon $r_d=r_s(z_d)$ at the drag epoch $z_d$ is
\begin{equation}
\label{eq:r_d}
r_s(z_d)=\int_{z_d}^\infty \frac{c_s(z)}{H(z)}dz,
\end{equation}
the sound speed in the baryon-photon plasma is $c_s(z)=1/\sqrt{3[1+\bar{R_b}/(1+z)}]$,
$\bar{R_b}=3\Omega_b h^2/(4\times 2.469\times 10^{-5})$
and $h=H_0/(100\text{ km s}^{-1}\text{Mpc}^{-1})$.
Since DESI BAO measures $D_M(z)/r_d$ and
$D_H(z)/r_d$ with an undetermined value of the sound horizon $r_d$, this limits the direct use of the data.
However, by deriving the AP parameter $F_{AP}=D_M(z)/D_H(z)$ from the DESI BAO measurements,
the dependence on $r_d$ is eliminated.
Therefore, we will use the derived AP parameter to perform the null test.

\begin{figure}[H]
\centering
\includegraphics[width=0.35\textwidth]{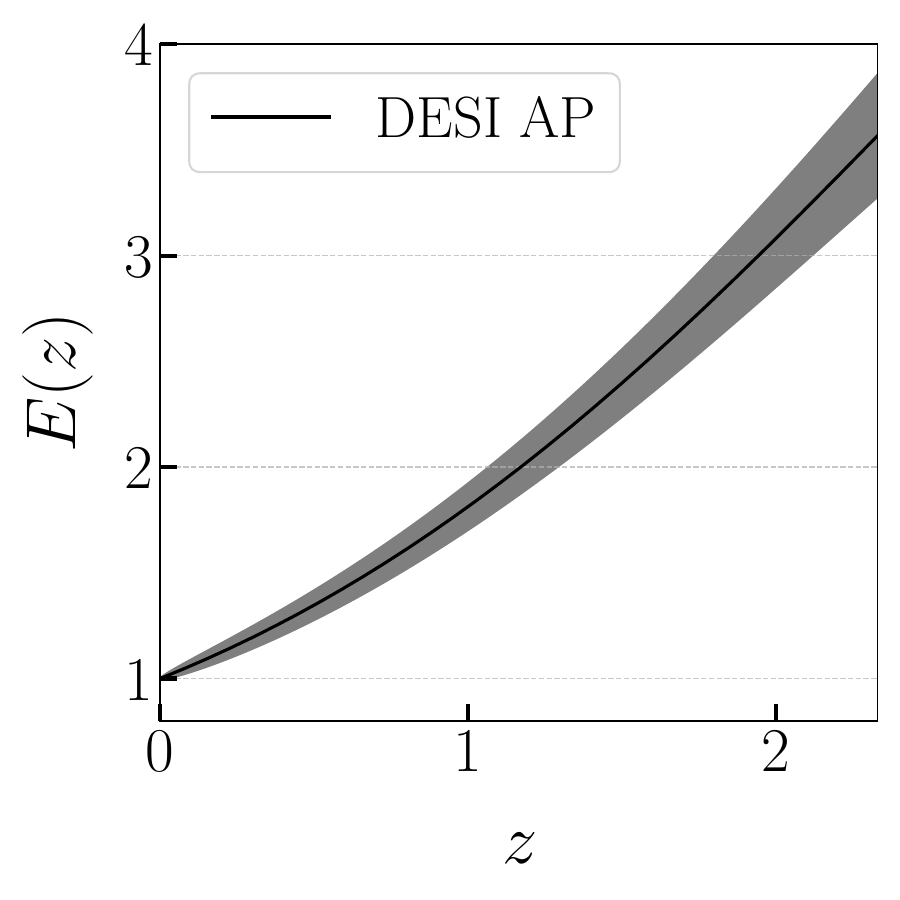}
\caption{The reconstructed $E(z)$ along with the $1\sigma$ confidence level from DESI BAO AP data.
The solid line is for the reconstructed mean and the shaded region is for the $1\sigma$ confidence level. }
\label{fig:fap_ez}
\end{figure}

In a flat universe, the AP parameter $F_{AP}$ is
\begin{equation}
\label{eq:fap_ez}
F_{AP}=E(z)\int_0^z\frac{1}{E(x)}dx=E(z)D(z),
\end{equation}
so
\begin{equation}
E'(z)=E(z)\frac{F'_{AP}(z)-1}{F_{AP}(z)}.
\end{equation}
For a spatially flat universe, we can reconstruct $E(z)$ and $E'(z)$ from DESI BAO AP data and then use the reconstructed $E(z)$ to conduct null tests.
Using Eq. \eqref{eq:fap_ez}, we reconstruct $E(z)$ from the DESI BAO AP data using the GP with the GaPP package \cite{Seikel:2013fda},
and the result is shown in Fig. \ref{fig:fap_ez}.
For the covariance function, we choose the widely-used squared-exponential kernel, which is  infinitely differentiable.
Even though the reconstruction of the Hubble parameter from $H(z)$ data shows some differences between the stationary and non-stationary kernels, especially at high redshifts \cite{Johnson:2025blf},
the results from GP reconstruction are usually not sensitive to the choice of kernels and hyperparameters \cite{Yu:2017iju,Dinda:2024ktd,Sabogal:2024qxs}.

On the other hand, we can reconstruct $D_M(z)$ and $D(z)$ from DESI BAO $D_M/r_d$ data to perfrom the $Ok$ and $Lz$ diagnostics.

\subsection{$Ok$ diagnostic}
To reconstruct $\mathcal{O}k(z)$ from DESI BAO data, we need to know both $E(z)$ and $D(z)$.
However, the $E(z)$ reconstructed from DESI BAO AP data in Fig. \ref{fig:fap_ez} assumes that $\Omega_{k0}=0$,
so we cannot apply the $Ok$ diagnostic to test the flatness of the spatial geometry and the cosmological principle.
Instead, we reconstruct $\mathcal{O}k(z)$ from DESI BAO data to evaluate the consistency of the assumption of a flat geometry.

\begin{figure}[H]
\centering
\includegraphics[width=0.35\textwidth]{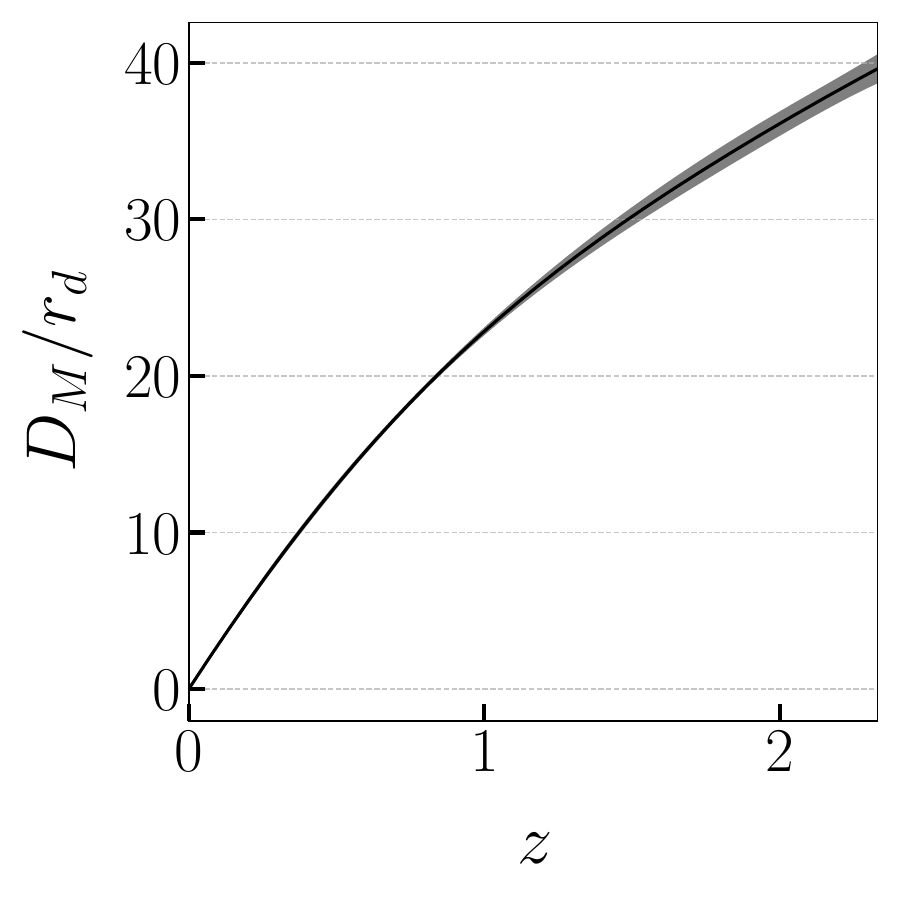}
\caption{The reconstructed $D_M/r_d$ along with the $1\sigma$ confidence level from DESI BAO $D_M/r_d$ data. The solid line is for the reconstructed mean and the shaded region is for the $1\sigma$ confidence level. }
\label{fig:dz}
\end{figure}

\begin{figure}[H]
\centering
\includegraphics[width=0.35\textwidth]{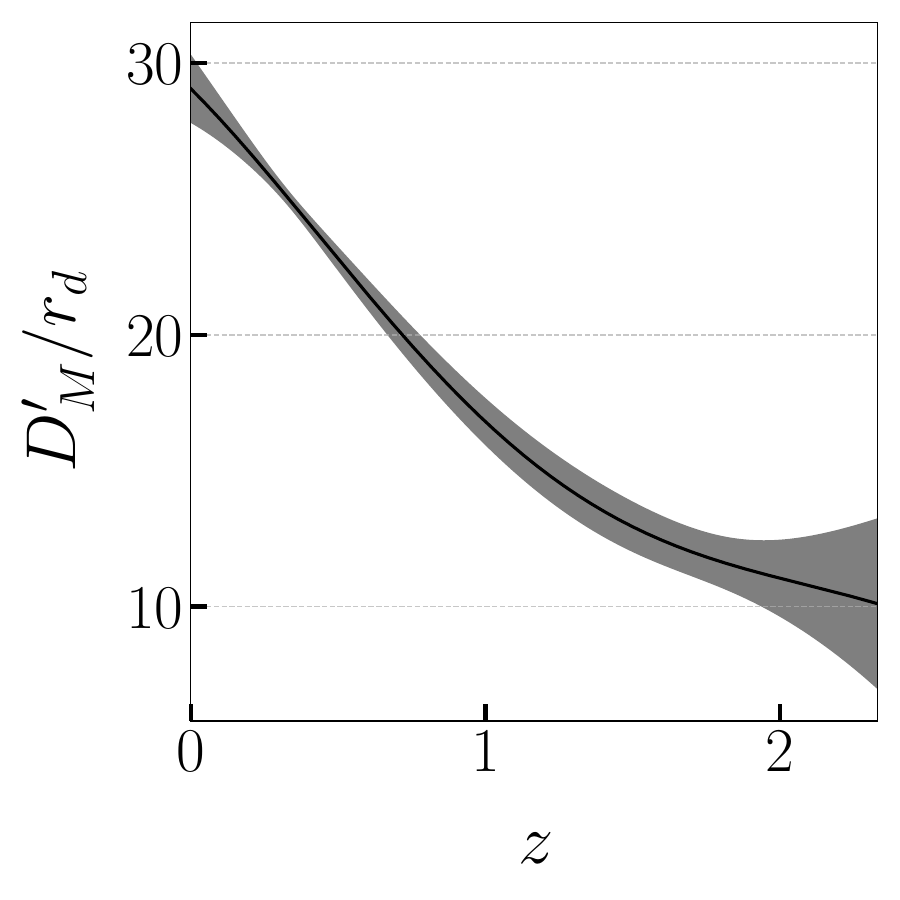}

\caption{The reconstructed $D_M'/r_d$ along with the $1\sigma$ confidence level from DESI BAO $D_M/r_d$ data. The solid line is for the reconstructed mean and the shaded region is for the $1\sigma$ confidence level. }
\label{fig3}
\end{figure}

\begin{figure}[H]
\centering
\includegraphics[width=0.45\textwidth]{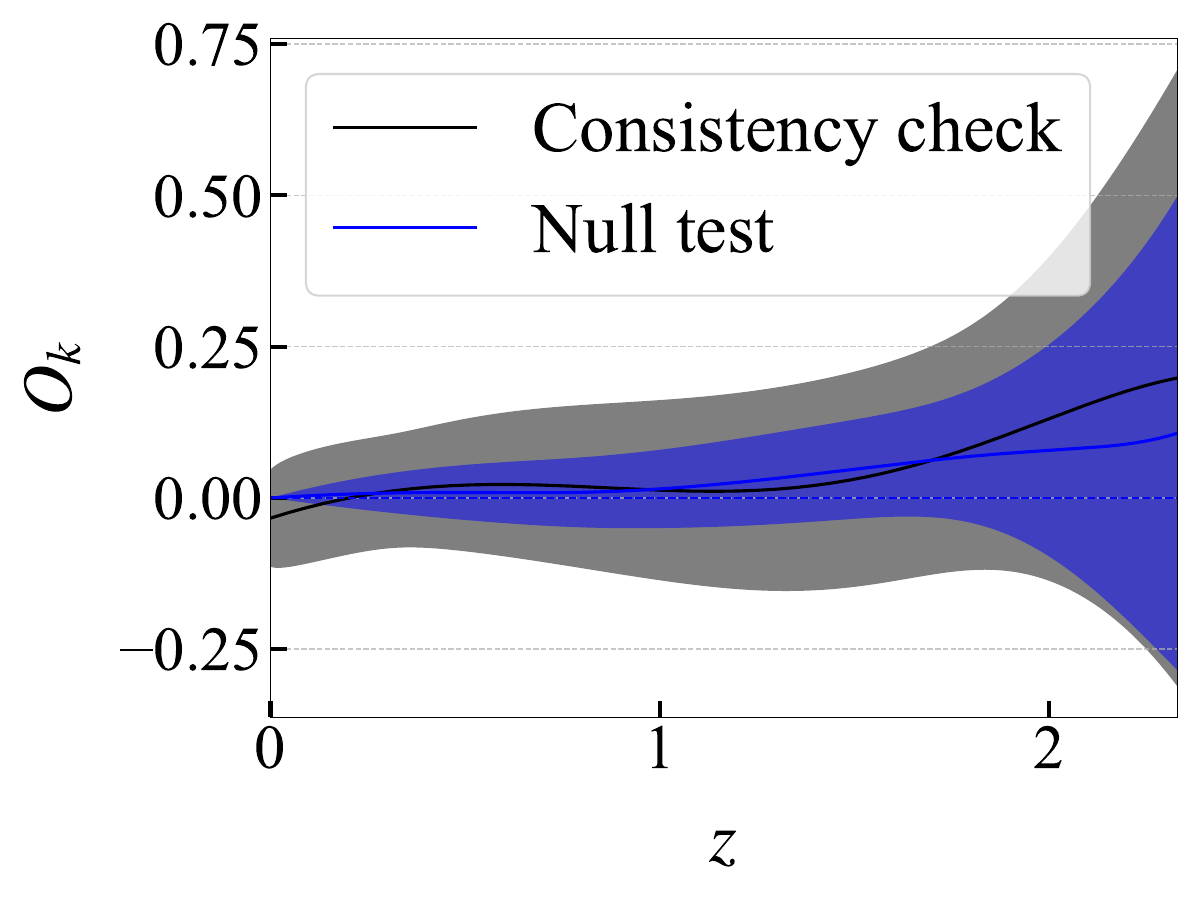}
\caption{The reconstructed $\mathcal{O}_k$ along with the $1\sigma$ confidence level from DESI BAO.
The solid lines are for the reconstructed mean and the shaded regions are for the $1\sigma$ confidence levels.
The black line and the gray region are the results reconstructed from $E(z)$ and $D_M/r_d$, we label them as consistency check. The blue line and region are the results reconstructed with Eq. \eqref{eq:fap_okz}, we label them as null test.}
\label{fig:fap_Ok}
\end{figure}

From the DESI BAO data, we reconstruct $D_M(z)/r_d$ and $D'_M(z)/r_d$ using GP as shown in Figs. \ref{fig:dz} and \ref{fig3}.
So if we know the value of $r_d h$, we can derive $D(z)$ and $D'(z)$ from the reconstructed $D_M/r_d$ and $D'_M(z)/r_d$.
In Ref. \cite{Lu:2024hvv}, the model-independent value $r_dh=99.8\pm3.1$ Mpc was obtained from
the DESI BAO $F_{AP}$ and $D_H/r_d$ data.
Note that the reconstruction of $E(z)$ from $F_{AP}$ assumes spatial flatness,
the derived value of $r_d h$ also depends on the assumption of $\Omega_{k0}=0$.
Due to the dependence of $r_d h$ on spatial flatness,
the reconstructed $D(z)$ somehow assumes $\Omega_{k0}=0$ too,
so this reconstruction from DESI BAO data does not serve as a null test for zero curvature and the cosmological principle.
By substituting the model-independent value of $r_dh$ into the reconstructed $D_M(z)/r_d$ and $D'_M(z)/r_d$,
we get $D(z)$ and $D'(z)$.
Combining the reconstructed results of $E(z)$ and $D(z)$,
we calculate $\mathcal{O}_k(z)$ with Eq. \eqref{ok1},
and the result is shown in Fig. \ref{fig:fap_Ok}.
From Fig. \ref{fig:fap_Ok}, we see that $\Omega_{k0}=0$ is consistent with DESI BAO data.

In the previous studies on the $Ok$ diagnostic \cite{Cai:2015pia,Wei:2016xti,Yu:2017iju,Marra:2017pst}, the authors reconstruct $E(z)$ from the $H(z)$ data, and reconstruct $D(z)$ either from BAO or SNe Ia data,
so they need to assume a value of the Hubble constant $H_0$ or the sound horizon $r_d$,
resulting in a loss of model independence in the conclusions.
As discussed above, if we follow the conventional method of reconstructing $E(z)$ and $D(z)$ from the DESI BAO data $D_M/r_d$ and $D_H/r_d$,
we must first assume spatial flatness and a specific value of $r_d$.
To circumvent these issues, we propose a novel approach that utilizes the DESI BAO AP data to conduct a null test on spatial curvature.

For a flat universe, Eq. \eqref{eq:fap_ez} tells us that
\begin{equation}
\label{eq:fap_okz}
\mathcal{O}_k=F_{AP}\frac{D'}{D}-1=F_{AP}\frac{D_M'/r_d}{D_M/r_d}-1.
\end{equation}
Therefore, we can use Eq. \eqref{eq:fap_okz} to test the consistency between a spatially flat universe and DESI BAO data and avoid the problem of the dependence on the value of $r_d$.
The non-parametric reconstructions of $F_{AP}$, $D_M/r_d$ and $D'_M/r_d$ from DESI BAO data are independent of cosmological models,
so the combination of DESI BAO $F_{AP}$ and $D_M/r_d$ data can null test the spatial flatness and the cosmological principle.
We show the reconstructed $\mathcal{O}_k$ along with the $1\sigma$ confidence level from DESI BAO data in Fig. \ref{fig:fap_Ok}.
From Fig. \ref{fig:fap_Ok}, we see that $\mathcal{O}_k=0$ is consistent with DESI BAO  data at $1\sigma$ confidence level.

\subsection{$Om$ diagnostic}
\begin{figure}[H]
\centering
\includegraphics[width=0.35\textwidth]{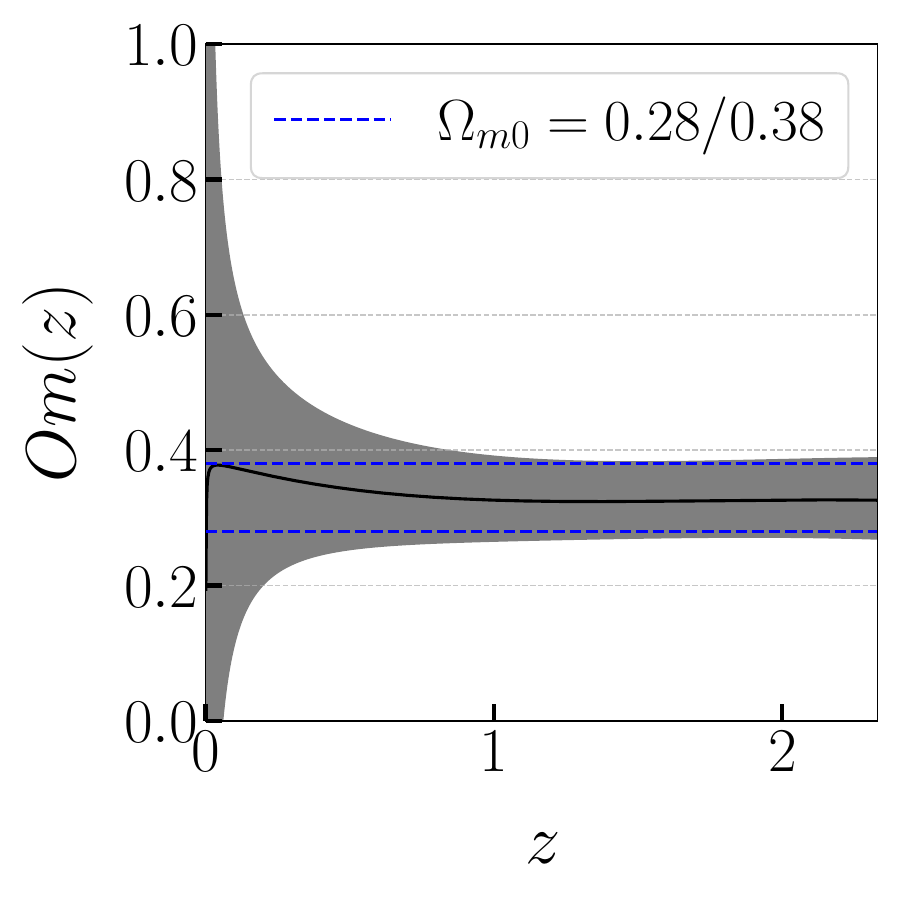}
\caption{The reconstructed $Om(z)$ along with the $1\sigma$ confidence level from DESI BAO AP.
The solid line is for the reconstructed mean and the shaded region is for the $1\sigma$ confidence level. The blue dashed lines correspond to the results for the flat $\Lambda$CDM model with $\Omega_{m0}=0.28$ and $0.38$, respectively.}
\label{fig:fap_om}
\end{figure}

Now we test the flat $\Lambda$CDM model with the $Om$ diagnostic.
Substituting the reconstructed $E(z)$ from DESI BAO AP data into Eq. \eqref{om1}, we get $Om(z)$ as shown in Fig. \ref{fig:fap_om}.
From Fig. \ref{fig:fap_om}, we see that flat $\Lambda$CDM models with $0.28\le \Omega_{m0}\le 0.38$ are consistent with the DESI BAO data at the $1\sigma$ confidence level.

\subsection{$Lz$ test}
\begin{figure}[H]
\centering
\includegraphics[width=0.35\textwidth]{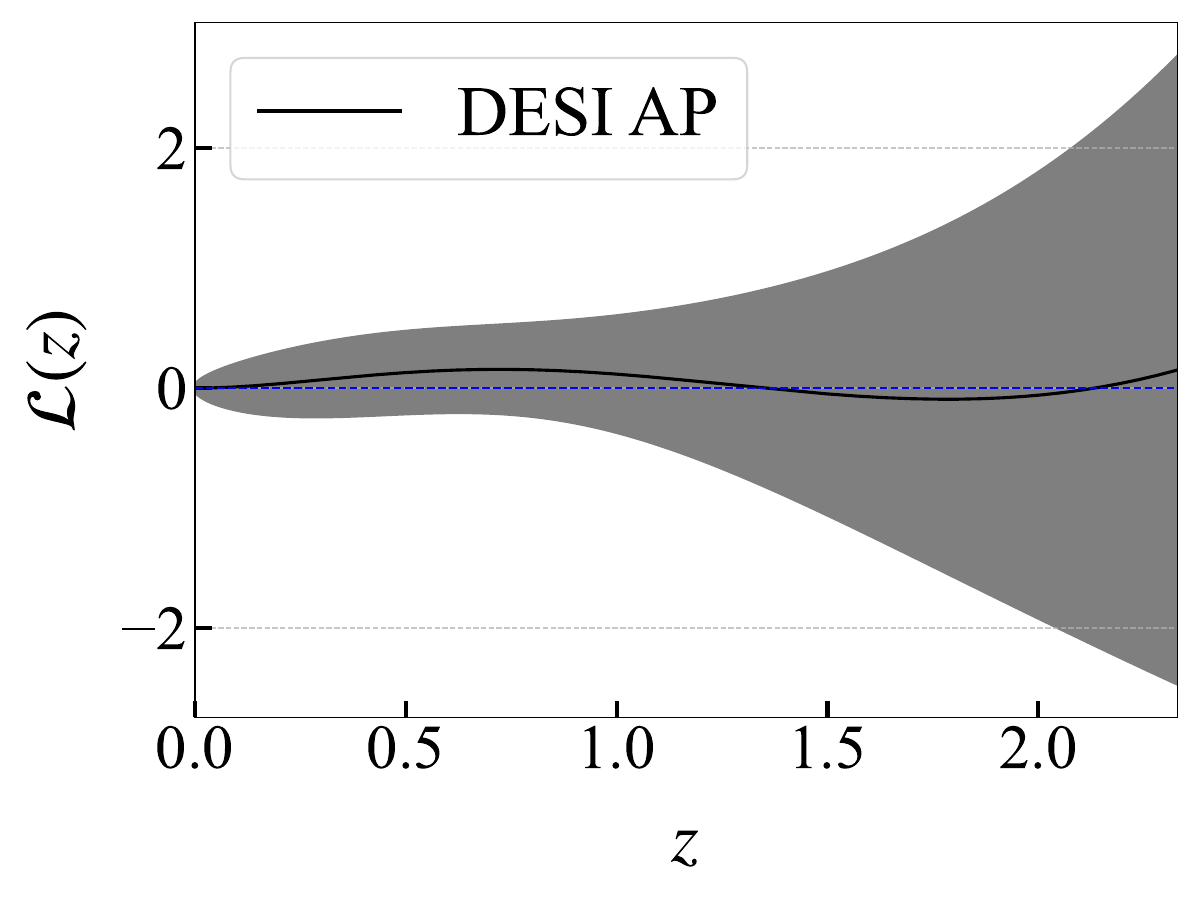}
\caption{The reconstructed $\mathcal{L}(z)$ along with the $1\sigma$ confidence level from DESI BAO AP.
The solid line is for the reconstructed mean and the shaded region is for the $1\sigma$ confidence level. }
\label{fig:fap_Lz}
\end{figure}

Since for a spatially flat universe, $D'(z)=1/E(z)$, so we use the reconstructed $E(z)$ and $E'(z)$ from the DESI BAO AP data to perform the $Lz$ test.
Substituting the reconstructed $D'(z)$ and $D''(z)$ from the reconstructed $E(z)$ shown in Fig. \ref{fig:fap_ez} into Eq. \eqref{lz1},
we get the $Lz$ null test and the result is shown in Fig. \ref{fig:fap_Lz}.
The result shows that the flat $\Lambda$CDM model is consistent with the DESI BAO data at the $1\sigma$ confidence level.

\section{CCH and SNe Ia data}
\label{snh_null}
\begin{figure}[H]
\centering
\includegraphics[width=0.45\textwidth]{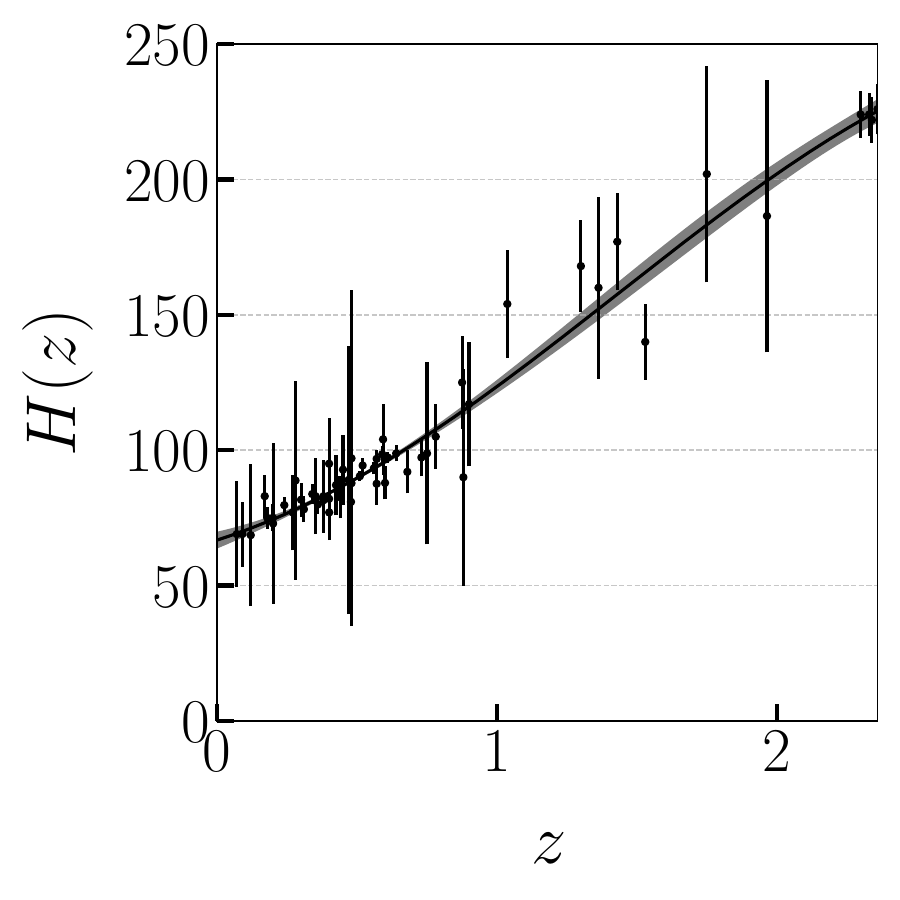}
\caption{The reconstructed Hubble parameter $H(z)$ (the solid line) along with the $1\sigma$ confidence level (the shaded region) from the CCH data. The original data points along with their $1\sigma$ confidence levels are also shown.}
\label{fig:hzrec}
\end{figure}

In this section, we use the CCH and SNe Ia data to perform null tests.
We reconstruct the Hubble parameter $H(z)$ using the $H(z)$ data compiled in Ref. \cite{Gadbail:2024rpp}.
The $H(z)$ data labelled as H, covers the redshits $0.07<z<2.36$, including 32 data points obtained with the cosmic chronometer
 (CCH) method \cite{Jimenez:2001gg,Simon:2004tf,Stern:2009ep,Zhang:2012mp,Moresco:2012jh,Moresco:2015cya,Moresco:2016mzx,Ratsimbazafy:2017vga,Borghi:2021rft},
and 26 data points  derived from radial BAO observations \cite{Gaztanaga:2008xz,Chuang:2012qt,Blake:2012pj,BOSS:2012gof,BOSS:2013rlg,Oka:2013cba,BOSS:2013igd,BOSS:2014hwf,BOSS:2016zkm,BOSS:2016wmc,BOSS:2017fdr}.
The reconstructed $H(z)$ using the GP is shown in Fig. \ref{fig:hzrec}.
From Fig. \ref{fig:hzrec}, we see that the Hubble constant is $H_0=66.78\pm 3.11$ km/s/Mpc.

\subsection{$Ok$ diagnostic}
\begin{figure}[H]
\centering
\includegraphics[width=0.35\textwidth]{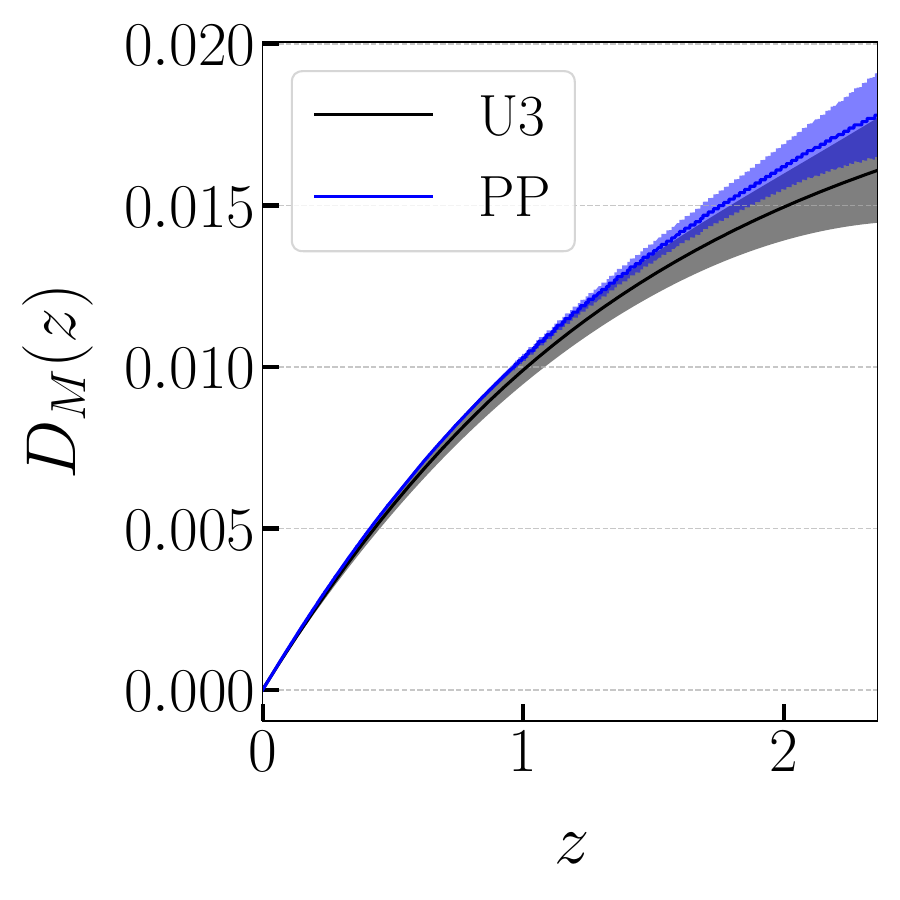}

\caption{The reconstructed $D_M$ along with the $1\sigma$ confidence levels from the Pantheon Plus (PP) and Union3 (U3) SNe Ia data. The solid lines are for the reconstructed mean and the shaded regions are for the $1\sigma$ confidence level. }
\label{fig:dmorg}
\end{figure}

\begin{figure}[H]
\centering
\includegraphics[width=0.35\textwidth]{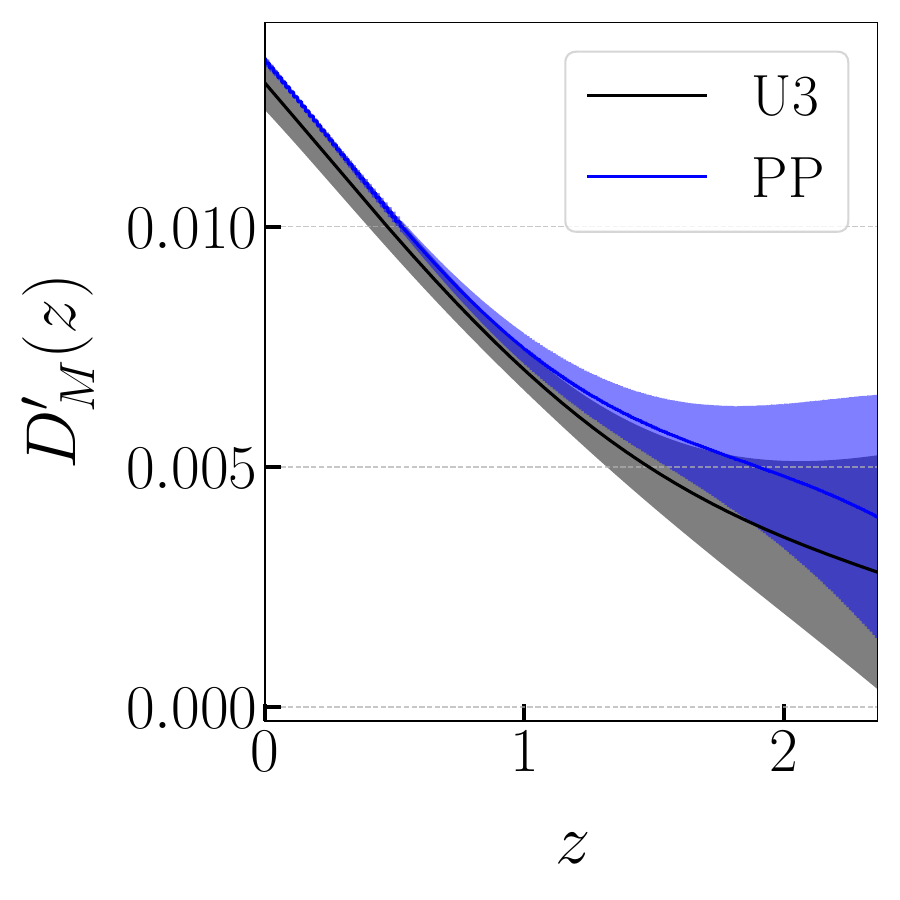}

\caption{The reconstructed $D_M'$ along with the $1\sigma$ confidence levels from the Pantheon Plus (PP) and Union3 (U3) SNe Ia data. The solid lines are for the reconstructed mean and the shaded regions are for the $1\sigma$ confidence level. }
\label{fig9}
\end{figure}

In this subsection, we combine the CCH and SNe Ia data to conduct $Ok$ diagnostic.
To avoid the dependence on $H_0$ as discussed in the previous section,
we plug $H(z)$ and $D_M'(z)$ data into Eq. \eqref{ok1} to perform the $Ok$ diagnostic.
We reconstruct the $D_M(z)$ from SNe Ia data.
We use two different SNe Ia data:
%1829 SNe Ia compiled by DES \cite{DES:2024jxu},
the Union3 compilation of 2087 SNe Ia \cite{Rubin:2023ovl}, and the Pantheon Plus sample of 1550 spectroscopically confirmed SNe Ia \cite{Scolnic:2021amr}.
We label
%the DES SNe Ia dataset as D5,
the Union3 SNe Ia dataset as U3,
and the Pantheon Plus SNe Ia dataset as PP.
The distance modulus $\mu$ is related with the luminosity distance by the following equation,
\begin{equation}
\label{distmod1}
    \mu = 5\log_{10}(d_L/\text{Mpc})+25,
\end{equation}
so we can reconstruct the transverse comoving distance $D_M(z)=d_L(z)/(1+z)$ from the measurements of the distance modulus $\mu$ by SNe Ia observations,
and the results of $D_M(z)$ and $D'_M(z)$ are shown in Figs. \ref{fig:dmorg} and \ref{fig9}, respectively.

\begin{figure}[H]
\centering
\includegraphics[width=0.35\textwidth]{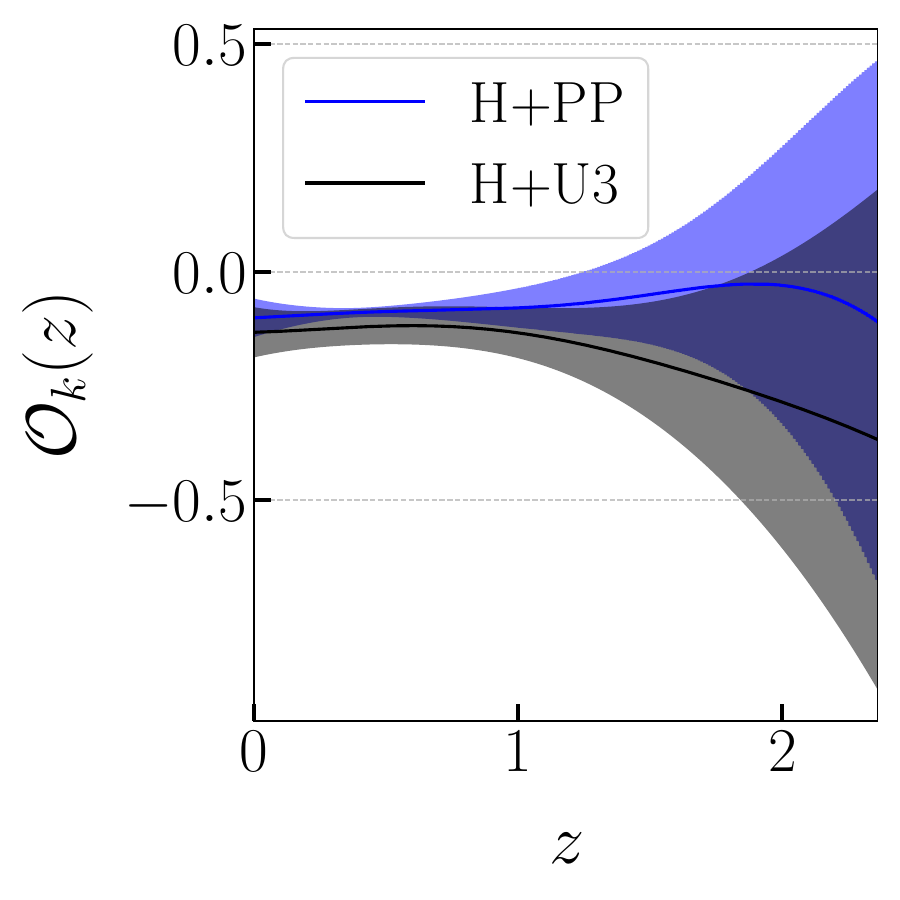}

\caption{The reconstructed $\mathcal{O}_k$ along with the $1\sigma$ confidence level from the combined CCH and SNe Ia data. The solid lines are for the reconstructed mean and the shaded regions are for the $1\sigma$ confidence level.}
\label{fig:okz}
\end{figure}

\begin{figure}[H]
\centering
\includegraphics[width=0.35\textwidth]{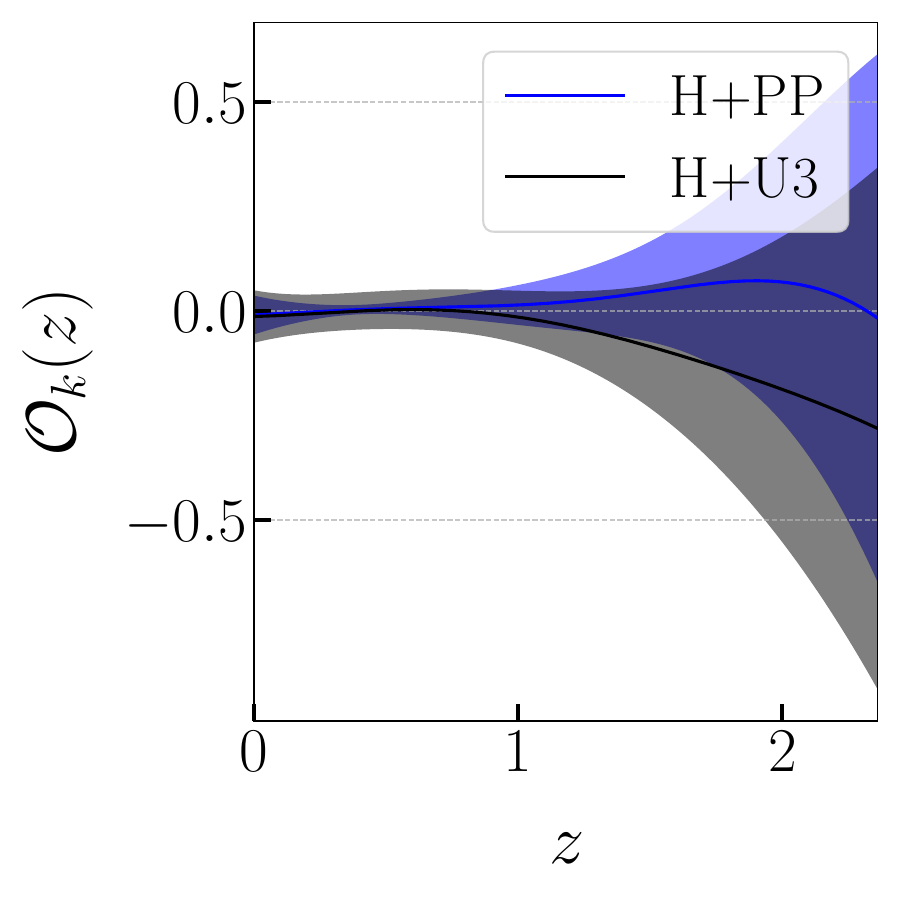}

\caption{The reconstructed $\mathcal{O}_k$ along with the $1\sigma$ confidence level from the combined CCH and SNe Ia data. The solid lines are for the reconstructed mean and the shaded regions are for the $1\sigma$ confidence level. An additional absolute magnitude of $-0.28$ and $-0.21$ was added to the Union3 (U3) and Pantheon Plus (PP) SNe Ia data, respectively.}
\label{fig11}
\end{figure}

Combining the reconstructed $H(z)$ and $D'_M(z)$, we get the reconstructed $\mathcal{O}_k(z)$ and the result is shown in Fig. \ref{fig:okz}.
From Fig. \ref{fig:okz}, we see that a constant but nonzero $\mathcal{O}_k(z)$ is consistent with combined CCH and SNe Ia data at the $1\sigma$ level,
indicating that the observational data imply the Universe is spatially flat and that the cosmological principle is supported by observations.
Note that if $\Omega_{k0}\neq 0$, then $\mathcal{O}_k(z)$ cannot be a constant in general,
so a constant $\mathcal{O}_k(z)$ must mean a spatial flatness.
However, zero curvature means $\mathcal{O}_k(z)=0$.
Therefore, a constant but nonzero $\mathcal{O}_k(z)$ may happen if the Hubble constant $H_0$ inferred from the $H(z)$ data is different from that derived from the SNe Ia data.
Because the SNe Ia data suffers the issue of zero-point calibration, or there exists the degeneracy between $H_0$ and the absolute magnitude of an SNe Ia,
the value of $H_0$ inferred from SNe Ia data may not reflect the true Hubble constant,  resulting in a constant but nonzero $\mathcal{O}_k(z)$.
Therefore, the nonzero $\mathcal{O}_k(z)$ reconstructed from the combined data may suggest a need to adjust the absolute magnitude.
By adding an additional absolute magnitude of $-0.28$ and $-0.21$ to U3 and PP data, respectively, we reconstruct $\mathcal{O}_k(z)$ again, and the results are shown in Fig. \ref{fig11}.
From Fig. \ref{fig11}, we see that $\mathcal{O}_k(z)=0$ is consistent with the combined CCH and SNe Ia data at the $1\sigma$ level after we adjust the absolute magnitude of SNe Ia data.

Adding an additional absolute magnitude of $-0.28$ to U3 data and fitting the flat $\Lambda$CDM model, we get $\Omega_{m0}=0.357\pm0.027$ and $H_0=63.7\pm2.6$ km/s/Mpc.
Adding an additional absolute magnitude of $-0.21$ to PP data and fitting the flat $\Lambda$CDM model, we get $\Omega_{m0}=0.332\pm0.018$ and $H_0=66.49\pm0.21$ km/s/Mpc.
These results may suggest that the Hubble tension comes from the zero-point calibration of SNe Ia data.

\subsection{$Om$ diagnostics}
\begin{figure}[H]
\centering
\includegraphics[width=0.35\textwidth]{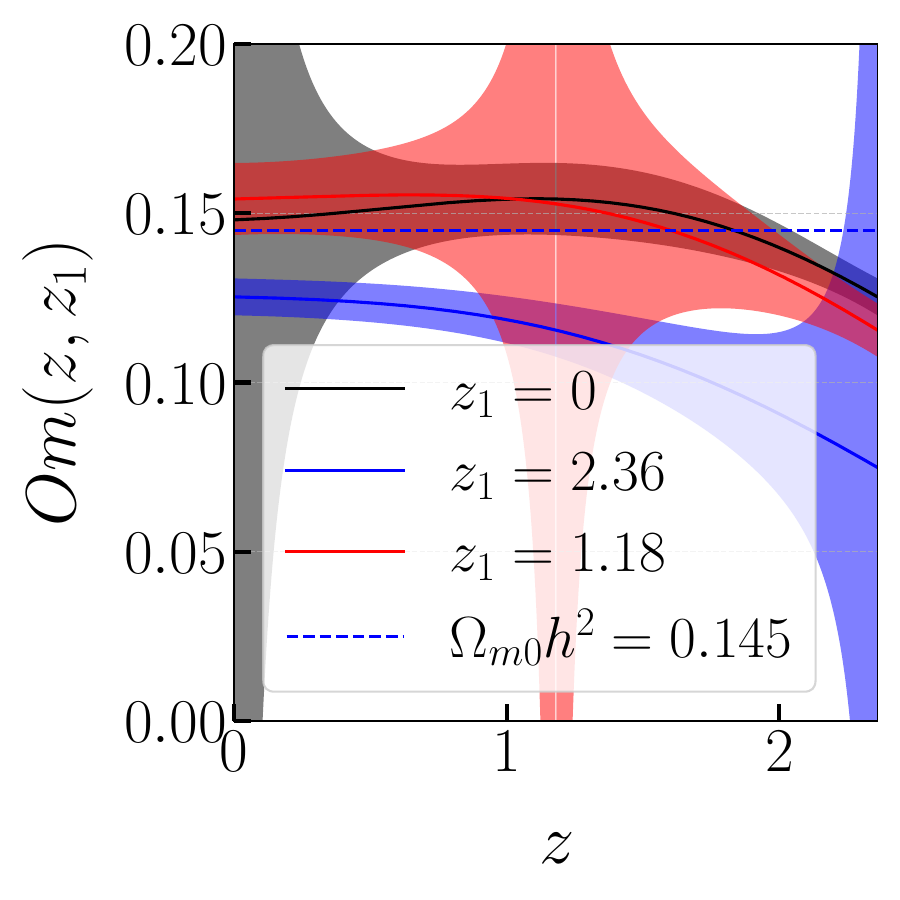}
\caption{The reconstructed $Om(z,z_1)$ along with the $1\sigma$ confidence level from the CCH data.
The solid lines are for the reconstructed mean and the shaded regions are for the $1\sigma$ confidence levels. The blue dashed line corresponds to the result for the flat $\Lambda$CDM model with $\Omega_{m0}h^2=0.145$.}
\label{fig12}
\end{figure}

In this section, we apply the two-point $Om$ diagnostic Eq. \eqref{om2a} to conduct the null test on the flat $\Lambda$CDM model using the CCH data.
Plugging the reconstructed $H(z)$ shown in Fig. \ref{fig:hzrec} to Eq. \eqref{om2a} and choosing different values for $z_1$,
we get the result $Om(z,z_1)$ as shown in Fig. \ref{fig12}.
From Fig. \ref{fig12}, we see that the flat $\Lambda$CDM is inconsistent with the $H(z)$ data at the $1\sigma$ level,
and the deviation is significant at high redshift $\gtrsim 1.5$.
Surprisingly, the reconstructed $Om(z,z_1)$ with $z_1=2.36$ differs from those for $z_1=0$ and $z_1=1.18$.
Since $H(z)$ data we used come from different methods, in particular, the data points with $z>2$ were derived from methods different from CCH,
this is perhaps the reason for the inconsistency.
Anyhow, the inconsistency deserves further investigation.
This issue with the CCH data may also impact the calibration of the absolute magnitude of SNe Ia, we will study it in future work.

\section{Conclusion}
\label{conclusion}
To test the spatial flatness and the cosmological principle using the $Ok$ diagnostic,
we propose two different methods.
The first non-parametric method involves reconstructing $F_{AP}$, $D_M/r_d$ and $D'_M/r_d$ from DESI BAO data, then calculating $\mathcal{O}_k$ with Eq. \eqref{eq:fap_okz} to perform the null test.
This novel method avoids the issue of the value of the Hubble constant suffered from the reconstruction of $E(z)$ and $D(z)$ commonly used in the literature.
The second non-parametric method entails reconstructing  $H(z)$ from the CCH data and $D_M(z)$ from SNe Ia data, followed by calculating $\mathcal{O}_k$ with Eq. \eqref{ok1} to perform the null test.
The results show that a spatially flat universe and the cosmological principle are consistent with both DESI BAO data and the combination of CCH and SNe Ia data.
It is interesting to note that values of $-0.28$ and $-0.21$ should be added to the absolute magnitude of SNe Ia for the Union3 and Pantheon Plus SNe Ia data, respectively, to align the CCH data with the SNe Ia data.

For DESI BAO data, we also derive the dimensionless variable $E(z)$ from the reconstructed $F_{AP}$ by assuming a spatially flat universe.
Using the model-independent value $r_dh=99.8\pm3.1$ Mpc, we obtain $D(z)$ and $D'(z)$ from the reconstructed $D_M/r_d$ and $D'_M/r_d$,
checking the consistency of a flat universe with the DESI BAO data.
We find that $\Omega_{k0}$ aligns with DESI BAO data.
Note that the reconstruction of $E(z)$ from $F_{AP}$ assumes spatial flatness,
and the derived value of $r_d h$ also relies  on the assumption of $\Omega_{k0}=0$,
making this test a consistency check rather than a null test.
With the reconstructed $E(z)$ and $E'(z)$, we also perform the $Om$ diagnostic and $Lz$ test,
finding that the flat $\Lambda$CDM model is consistent with DESI BAO data.

For the reconstructed $H(z)$ from CCH data, we also conduct the $Om(z_2,z_1)$ diagnostic by choosing the first point from the beginning, middle and end of the CCH data.
In particular, we select $z_1=0$, 1.18 and 2.36, and we find that the flat $\Lambda$CDM model is consistent with CCH data up to the redshift $z\lesssim 1.5$ for $z_1=0$ and $z_1=1.18$.
Surprisingly, the reconstructed $Om(z,z_1)$ with $z_1=2.36$ differs significantly from those with $z_1=0$ and $z_1=1.18$.
The behavior of $H(z)$ data is unexpected and requires further investigation.

In conclusion, there is no evidence of deviation from the flat $\Lambda$CDM model, nor is there any indication of dynamical dark energy found in the observational data. Since we employ a non-parametric reconstruction method, this conclusion remains robust and agnostic to any cosmological model and gravitational theory.

\begin{acknowledgments}
This work is supported in part by the National Key Research and Development Program of China under Grant No. 2020YFC2201504, the National Natural Science Foundation of China under Grant No. 12175184 and the Chongqing Natural Science Foundation under Grant No. CSTB2022NSCQ-MSX1324.
\end{acknowledgments}

% \bibliographystyle{scichina}
% \bibliographystyle{physlett}
% \bibliography{gp_null_test}

\end{document}